\documentclass[12pt]{iopart}  
\usepackage{graphicx} 

\begin{document}

\maketitle
 
\title{Minimal Model for the Topology of the Critical State in Hard Superconductors}

\author{A. Bad\'{\i}a and C. L\'opez$^{*}$}

\address{
Dep. de F\'{\i}sica de la Materia Condensada-I.C.M.A., Univ. de Zaragoza, SPAIN
\\$^{*}$Dep. de Matem\'atica Aplicada, Univ. de Zaragoza, SPAIN
}

\begin{abstract}

The critical state problem in type-II superconductivity is described
theoretically by a direct optimization method, which allows a straightforward treatment for non idealized geometries. 

Based on Faraday's law and the principle of minimum entropy production, the magnetic history is built up just by a constrained minimization of the field changes along the process. Constraints are in the form $\vec{J}\in\Delta$, with $\vec{J}$ the electric current density and $\Delta$ some bounded set. This incorporates the vortex pinning and interaction phenomena and may be used for the modelling of anisotropy, inhomogeneities and flux cutting interactions.

In this work, our variational statement is posed on the finite element discretization and provides a minimal tool for investigating the effects of the sample's topology on the field penetration patterns. Simulations of (i) the contraction and splitting of boundaries between the flux free and penetrated regions, (ii) the effect of granularity on the superconducting properties, (iii) the influence of defects, and (iv) surface curvature phenomena are presented.

\end{abstract}

\section{INTRODUCTION}

Magnetization measurements related to the flux penetration and exit in type-II superconductors have been a benchmark against which various models of the interaction between flux quanta and the underlying pinning structure are tested\cite{campbellevetts}. The critical state theory\cite{bean} provides a quite intuitive and mathematically simple framework for analyzing the above mentioned macroscopic experiments. In brief, one is just led to solve Amp\`ere's law in the form $|\vec{\nabla}\times\vec{H}|=J_{c}$, with $J_{c}$ a material dependent parameter, the so-called critical current density. This becomes a trivial problem for idealized geometries (circular cylinders and slabs in longitudinal applied field), but may be really tough when one tries to simulate real world problems. This includes to allow for arbitrary sample shapes, spatial inhomogeneities, and more general restrictions on the current density than the original critical state equation $J=\pm J_{c},0$ for one-dimensional systems. Several numerical methods have been developed, which partially include the previous requirements, and  may be classified in two groups.

On the one side, some models deal with the full electromagnetic problem which arises when flux is changed within the superconductor, by means of a definite current-voltage law (typically in the form $E=E_{c}(J/J_{c})^{n}$, $n$ being a phenomenological exponent)\cite{brandt,schuster,mayergoyz}. This kind of theory leads to diffusion equations for the electromagnetic fields, which have been solved by a number of techniques. They do never hold a true critical state, unless for the limit $n\to\infty$.

On the other hand, the general critical state theories avoid using intermediate electric fields and just concentrate on the metastable states which can be generated by applying external magnetic field changes\cite{prigozhin,badiaprlb}. The material properties are expressed in the form of the restriction for the current density vector $\vec{J}\in\Delta$, with $\Delta$ some bounded set, as dictated by the physics of the problem. To be specific, the system either allows dissipativeless current flow for $\vec{J}$ within the region $\Delta$, or undergoes a transition to a very high dissipation regime (infinite in the mathematical limit), which is instantly rejected when $\vec{J}\notin\Delta$. As it has been discussed elsewhere \cite{badiajltp}, the selection of $\Delta$ is made on the basis of the physical interactions between  vortices and the substrate pinning potential. Thus, a restriction on $\vec{J}$ locally perpendicular to the magnetic field ($J_{c\perp}$) relates to the pinning interaction of line vortices, and a restriction on the component of $\vec{J}$ parallel to the magnetic field ($J_{c\parallel}$), comes from intervortex tilt interactions. Also, material isotropy can be incorporated by choosing $\Delta$ as a circle, or anisotropy, for instance, selecting $\Delta$ in the form of an ellipse. 
Notice that the one-dimensional critical state hypothesis is included as the particular case $\Delta = [-J_{c},J_{c}]$. The actual solution $J=\pm J_{c},0$, which was brilliantly intuited by C. P. Bean\cite{bean}, arises from the variational interpretation of Faraday's law as a rule of minimum magnetic field changes.

Along the past years, minimum principle formulations for the magnetic behavior of superconductors have been exploited, allowing a simple interpretation of many experimentals facts\cite{prigozhin,badiaprlb,badiajltp,badiajap}. These include the analysis of crossed field and rotation field experiments, anisotropic responses and the influence of sample's geometry.

It should be emphasized that generalized variational methods are not merely a substitute to differential equation statements of classical physical problems. Additionally, they provide a method to generate laws which incorporate singularities, discontinuities or inequality constraints, which can be hardly dealt with by local differential statements. This is, for instance, the case of type-II superconductors, which can be described by the limit $n\to\infty$ in the power-law current-voltage characteristic (multivalued function). Such benefit was already treated in previous issues, where we basically discussed a number of physical observations related to the specific set $\Delta$ in the theory. As for the sample's geometry, only the slab geometry (one independent variable) was treated. In this paper, we exploit another advantage of the variational statement, i.e.: the ease to pose the problem for any geometry on the finite element discretization. This allows to simulate a variety of experimental observations related to the sample's topology.

The article is organized as follows. In Sec.\ref{secvar} we pose the physical problem of the critical state in two-dimensional geometry and give some general results for our variational statement. A numerical implementation based on the finite element discretization is presented in Sec.\ref{secnum}. This is then applied to several case studies in Sec.\ref{seccase}, i.e.: a cylinder with arbitrary cross section, a granular sample, and a sample with a point defect. Further applications and possible extensions of the theory are discussed in Sec.\ref{secdiscuss}.

\section{STATEMENT OF THE PROBLEM}
\label{secvar}

In first place, we will define the macroscopic fields involved in the critical state regime and their relation to measurable quantities. Our work has been developed within the assumption that the sample's response is dominated by the so-called {\em irreversible} contribution. This means that the equilibrium response of the flux line lattice in the absence of pinning centers as well as any possible surface currents are neglected. For global magnetization measurements in which the integrated magnetic moment of the sample is measured, this is a very good approximation within the region $H_{c1}\ll H \ll H_{c2}$, unless for weak pinning specimens (see Ref.\cite{krasnov} and references therein). Then, the coarse-grained electrodynamics if formulated in terms of
\begin{itemize}
\item[(i)] The {\em flux density} $\vec{B}(\vec{x})$ within the sample is the average of the microscopic field intensity $\vec{h}$, over a volume which encloses a big number of vortices.
\item[(ii)] The {\em magnetic field} $\vec{H}(\vec{x})$, in the absence of equilibrium magnetization, is linearly connected to $\vec{B}$ by $\vec{B}=\mu_{0}\vec{H}$.
\item[(iii)] The averaged current density may be calculated from $\vec{J}=\nabla\times\vec{H}$.
\end{itemize}
Finally, the connection to observable quantities is done by characterizing the external field sources and the sample's response.
\begin{itemize}
\item[(iv)] On neglecting finite size effects, the magnetic source ($\vec{H}_{S}$) enters as a boundary condition for the flux density at the surface of the sample. The tangential component is continuous $B_{T}(surface)=\mu_{0}H_{S,T}$.
\item[(v)] Then, the measured magnetic moment of the sample per unit volume is $\vec{M}\equiv\langle\vec{B}\rangle /\mu_{0}-\vec{H}_{S}=\langle\vec{H}\rangle -\vec{H}_{S}$. Here, the average concerns the whole volume of the superconductor.
\end{itemize}

\subsection{The critical state for long samples}

\begin{figure}
\centerline{
\includegraphics[height=3.in]{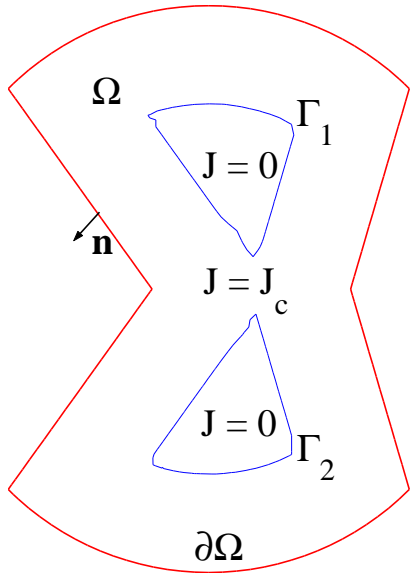}
}
\caption{Superconductor with disconnected subcritical regions established by application of an increasing magnetic field.}  
\label{fig:freebound}
\end{figure}

The magnetization of long specimens with arbitrary cross section was already treated by Campbell \& Evetts\cite{campbellevetts}. A number of general properties of the field penetration profiles were obtained on the basis of the critical state equation, which according to the upper guidelines may be expressed as
\begin{equation}
\label{eq:cece}
\left (\frac{\partial H}{\partial x}\right )^{2}+
\left (\frac{\partial H}{\partial y}\right )^{2}= J_{c}^{2}\quad
({\rm or}\quad 0) \; .
\end{equation}
However, this method is only valid for very simple geometries (f.i.: elliptic and rectangular sections) as, in other cases, the free boundary between the critical ($|\vec{J}|=J_{c}$) and subcritical ($\vec{J}=0$) regions is unknown, and at least, a reasonable guess is needed. This may become very tough as the boundary could even split, giving place to multiply connected topology. This is depicted in Fig.\ref{fig:freebound} where 
two subcritical regions appear for the initial magnetization process of a cylinder with concavities in the cross section. In general, a free boundary problem will arise between regions either affected or not by flux changes in a definite manipulation. Thus, the previous comment extends to the fully pentrated regime when the applied excitation is reversed.

Additionally, the underlying assumption $|\vec{J}|=J_{c}\;{\rm or}\; 0$ is clearly insufficient for describing the full phenomenological scenario. This requires a more general constitutive law of the kind $\vec{J}\in\Delta$\cite{badiajltp}.

Below, we describe the minimal model which accounts for all the aforementioned facts. Free boundaries are no longer a shortcoming, as they will arise from the solution of our statement, which easily applies for any shape of the sample. Special emphasis will be made on this fact through the application to samples with corners, voids, and granular properties. 

\subsection{Variational statement}

We proceed along the lines established in previous work\cite{badiaprlb}. Recall that Faraday's law in time discretized form is equivalent to minimizing the spatially integrated magnetic field changes over successive time layers, and that the superconducting properties enter in the form of the restriction $\vec{J}\in\Delta$.

Consider an infinite cylinder of arbitrary cross section subjected to an applied magnetic field parallel to the axis. By virtue of the symmetry, the internal field will also be parallel to the axis and continuous across the surface, and the induced current density will flow along the basal plane (XY in what follows). Thus, the Critical State problem may be formulated as the condition for evolutionary field penetration profiles:
\begin{equation}
\label{eqnvarpri}
{\tt minimize }\quad{\cal C}[H_{\rm n+1}(\vec{x})]
=\int_{\Omega}\! | H_{\rm n+1} -
H_{\rm n} |^{2}\quad {\rm for} \quad\vec{u}\in\Delta_{\perp} \; .
\end{equation}
In this equation, the following notation is used:
\begin{itemize}
\item[ ] $\Omega$ stands for the planar cross section region.
\item[ ] $H_{\rm n+1}$ is the inner Z component of the magnetic field at the time layer $(n+1)\delta t$.
\item[ ] $H_{\rm n}$ is the given previous profile.
\item[ ] $\vec{u}$ is the control variable. In this problem $\vec{u}={\bf grad} H=(-J_{y},J_{x})/J_{c}$, i.e.: the current density is measured in units of $J_{c}$.
\item[ ] $\Delta_{\perp}$ is the control region (restriction for $\vec{u}$, $90^{o}$ rotation of $\Delta$).
\end{itemize}
In former research, we have solved Eq.(\ref{eqnvarpri}) in terms of associated Hamilton differential equations within the Optimal Control theory. As we restricted our attention to the slab geometry in parallel field, a system of ordinary differential equations with prescribed boundary conditions had to be solved. Here, the situation is a bit more involved, as a partial differential equation problem arises. Nevertheless, the same underlying ideas may be used.

\subsection{Optimal control formulation}
On using the formalism of Classical Mechanics, we define a Hamiltonian density, containing the Lagrangian density to be minimized and associated auxiliary momenta
\[
{\cal H}({H}_{n+1}, {\vec u}, {\vec p}, \vec{x})\equiv{\vec p}\cdot {\vec
u}-{{1}\over {2}}|{H}_{n+1}-{H}_n|^2 \; .
\]
Denoting by ${H}_{n+1}^*(\vec{x})$, ${\vec p}^{\,*}(\vec{x})$ and ${\vec u}^{\,*}(\vec{x})$ the optimal solution functions (i.e. minimizing ${\cal C}$ and satisfying the control system), the optimal control equations are
\begin{equation}
{\bf grad}\;{H}_{n+1}^*=  \frac{\partial{\cal H}}
{\partial \vec{p}^{\,*}} = {\vec u} \; ,
\end{equation}
\begin{equation}
\label{eqnham2}
{{\rm div}\;{\vec{p}}^{\,*}} = - \frac{\partial {\cal H}}{\partial
{H}_{n+1}^*}=
{H}_{n+1}^*-{H}_n(\vec{x})
 \; .
\end{equation}
and the algebraic condition of maximality
\[
{\cal H} ({H}, {\vec u}^{\,*}, {\vec p}, x) \geq {\cal H} ( {H}, {\vec u}, {\vec p}, x)
\qquad \forall \;{\vec u} \in \Delta_{\perp} \; .
\]
It is apparent that this condition is satisfied for a control vector $\vec{u}$ with maximum projection over the associated momentum $\vec{p}$. This is completely equivalent to the problems with one independent variable. Within $\Omega$ one has 
${\vec u}^{\,*}={\vec{p}}^{\,*}/{p}^{\,*}$ (or 0), with the physical counterpart $|\vec{J}|=J_{c}$ (or 0). In general, one can still identify the critical state condition as 
\[
\vec{u}^{\,*}\in\partial_{0}\Delta_{\perp} \; ,
\]
with $\partial_{0}\Delta_{\perp}$ standing for the union of the boundary of the region $\Delta_{\perp}$ and the zero point ($\partial_{0}\Delta_{\perp} = \{\partial\Delta_{\perp}\}\bigcup\,\{\vec{0}\}$). 

In the case of the slab geometry, the main technical difficulty for solving the problem was the existence of a free (unknown) point dividing the region penetrated by flux changes and the region free of flux variations at a particular step. In the current case of interest (more than one independent variable) the free boundary is either a curve (2D systems) or even a surface (3D systems). Even more, as shown above (see Fig.\ref{fig:freebound}), the free boundary is quite commonly defined by a collection of disconnected parts. Nevertheless, as it was recognized by Prigozhin\cite{prigozhin}, the use of variational statements offers the advantage of applying direct numerical minimization methods in which the distinction between both regions is unnecessary. In this paper, we will also make use of this benefit for our formulation. We will introduce an optimization method, which operates over the whole superconducting region and from which the free boundary automatically arises. However, some further comments on the Hamiltonian formulation are convenient as they will provide the basis for the physical interpretation of our calculations. The boundary conditions to be satisfied by the canonical variables are a generalization of the case for one independent variable. On the one side, owing to the absence of demagnetizing effects ${H}_{n+1}^*$ is determined on the sample surface $\partial\Omega$ by continuity of the external applied field. 

On the other hand, the free boundary surrounding the regions unaffected by flux variations ($H_{n+1}(\Gamma)-H_{n}(\Gamma)=0$) is characterized, according to its undetermined position by
\[
{\vec{p}}^{\,*}(\Gamma)=0\; .
\]
Notice, that the general situation $\Gamma = \bigcup \Gamma_i$ is allowed.
\subsection{Physical interpretation of the momentum vector $\vec{p}$}
\begin{figure}
\centerline{
\includegraphics[height=3.0in]{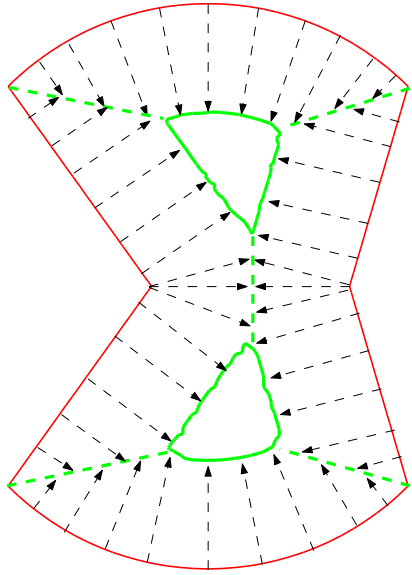}
}
\caption{Flux channels for the vortex penetration in a superconductor when magnetic field is increased.}  
\label{fig:paths}
\end{figure}

In this paragraph we show that the momentum $\vec{p}$, which has been introduced in the optimal control formulation as a Lagrange multiplier, has a clear physical interpretation. Recall that, in the absence of flux cutting phenomena, the continuity equation for the magnetic field may be written as 
\begin{equation}
\label{eq:continuity}
\frac{\partial H}{\partial t}+{\rm div}\vec{J}_{H}=0 \; ,
\end{equation}
where $\vec{J}_{H}$ stands for a magnetic field current density (not to be mistaken for the electric current density!). In simple words, flux variations within a region of the sample are due to the entry or exit of flux lines across its boundary. Now, comparing Eq.(\ref{eq:continuity}) and Eq.(\ref{eqnham2}) we conclude
\[
 \vec{p}\equiv -\vec{J}_{H}\,\delta t \;\; ,
\] 
with $\delta t$ the time step in our discretized model. Thus, solving for $\vec{p}$ allows to obtain the trajectories followed by vortices ({\em flux channels} hereafter) in the magnetization process. Recall that, isotropy provided, flux channels should be perpendicular to the sample's surface and finish at the free boundary $\Gamma$, which surrounds the regions unaffected by flux variations. This point has been illustrated in Fig.\ref{fig:paths}, where we display the integral lines for the vector $\vec{p}$ in the magnetization process of a superconductor with arbitary cross section. Dashed style has been used for the so-called {\em d-lines}\cite{schuster} (qualitative in our plot), which are never crossed by the incoming flux tubes. Continuous contours are used for the boundary between critical and subcritical regions. More details on the actual calculations leading to this plot are given later in the text.

\section{NUMERICAL SOLUTION}
\label{secnum}
In order to provide a wide application range of this work we describe a triangular Finite Element implementation of the numerical procedure for solving Eq.(\ref{eqnvarpri}). Although rectangular meshes are simple and intuitive, triangulation simplifies to deal with any geometry for the region $\Omega$ (sample's cross section). From the practical point of view, a big number of software facilities for the triangulation of any specified geometry can be found (and downloaded from the web). On the other hand, the fundamental aspects of the method may be found in numerous textbooks on the use of Finite Element Methods in electromagnetism (see f.i.\cite{bossavit,jin}). Here, we will just recall some notational aspects and a few relevant properties for our work.
%
%---------------------------------------------------
\subsection{Triangular mesh}
%---------------------------------------------------
%
As illustrated in Fig.\ref{fig:region}, the 2 dimensional region $\Omega$ will be {\em meshed} by a collection of triangles ${\cal T}$. A given triangle $T\in{\cal T}$ is defined by three nodes ($T=\{ k,l,m \}$). The set of all nodes is called $\cal N$.

Any physical property given by a function on $\Omega$ may be approximated in terms of the so-called {\em nodal functions} $\lambda_{k}(x,y)$, which are nothing but the {\em barycentric coordinates} of the point $(x,y)$\cite{bossavit}, referred to the triangle ${\rm T}$ to which this point belongs. On using the language of Classical Mechanics, ($\lambda_{k},\lambda_{l},\lambda_{m}$) are the values of normalized masses (thus, positive) which one should place at the vertices of ${\rm T}$, in order to get the center of mass at $(x,y)$. It is apparent that they provide an affine basis, in terms of which one can express the inner points of a given triangle
\[
\vec{x}=\lambda_{k}\vec{x}_{k}+\lambda_{l}\vec{x}_{l}+\lambda_{m}\vec{x}_{m} \; ,
\]
$\vec{x}_{k},\vec{x}_{l},\vec{x}_{m}$ standing for the coordinates of its vertices in some reference frame. The approximation of a function on the mesh is now apparent. For instance, the following expression may be used for the magnetic field at a given point
\begin{equation}
\label{expanH}
H(x,y)=\sum_{k\in{\rm T}} H_{k}\lambda_{k}(x,y) \; .
\end{equation}
Trivially, $H_{k}$ is nothing but the value of $H$ at the node $k$. Eq.(\ref{expanH}) will be the starting point for obtaining a matrix version of our variational statement (Eq.(\ref{eqnvarpri})).
\begin{figure}
\centerline{
}
\caption{Triangular mesh defined on a superconducting cylinder with arbitrary cross section $\Omega$.}  
\label{fig:region}
\end{figure}

Let us now review some relevant properties of the nodal functions. For a given triangle, with area $A_{T}$ one has
\[
\lambda_{\alpha}=\frac{1}{2A_{T}}(a_{\alpha}+b_{\alpha}x+c_{\alpha}y) \quad \alpha=k,l,m
\]
where
\begin{eqnarray}
a_{k}&=&x_{l}\,y_{m}-y_{l}\,x_{m}
\nonumber\\
b_{k}&=&y_{l}-y_{m}
\nonumber\\
c_{k}&=&x_{m}-x_{l}
\nonumber
\end{eqnarray}
and so on, by cyclic permutation. From these expressions one can verify the very important property that the nodal functions form a {\em partition of unity}, i.e.:
\[
\lambda_{k}(\vec{x})+\lambda_{l}(\vec{x})+\lambda_{m}(\vec{x})=1 \; .
\]
Based on all this, two fundamental results for our purposes immediately follow
\[
{\bf grad}\;\lambda_{\alpha}=(b_{\alpha} ,c_{\alpha})/2A_{T}  \quad {\rm on}\;\; T
\]
and
\begin{eqnarray}
\int_{T} \! \lambda_{\alpha}\lambda_{\beta}= 
\left\{
\begin{array}{ll}
A_{T}/6 & {\rm if} \quad \alpha =\beta \in T\\
A_{T}/12 & {\rm if} \quad \alpha \neq \beta \in T
\end{array}
\right.
\nonumber
\end{eqnarray}
%
%
%---------------------------------------------------
\subsection{Optimization problem}
%---------------------------------------------------
%
In this paragraph, we show that our variational statement (Eq.(\ref{eqnvarpri})) may be transformed into a nonlinear quadratic optimization problem by using the nodal functions expansion of the field (Eq.(\ref{expanH})) and the above displayed properties of the basis.

Thus, for the isotropic case, in which the critical current restriction may be written $|\vec{u}|=|{\rm{\bf grad}}\,H|\leq 1$, the magnetization problem consists of finding the set of mesh coordinates $\{ H_{n+1,k}\; , k\in {\cal N}\}$ for each time layer, so as to
\vspace{0.5cm}

{{\hspace{-.25cm}\fboxsep
1.ex\fbox{\parbox{0.9\textwidth}{ 
\begin{tabular}{rl}
${\tt Minimize}$ & $\quad\displaystyle{\sum_{T\in{\cal T}}\thinspace\sum_{\alpha ,\,\beta\in{T}}
(H_{\rm n+1,\alpha}-H_{\rm n,\alpha})I_{\alpha \beta}(H_{\rm n+1,\beta}-H_{\rm n,\beta})}$
\\
\\
${\tt for}$ & $\quad\displaystyle{\sum_{\alpha ,\,\beta\in{T}}
{H_{\rm n+1,\alpha}}G_{\alpha \beta}{H_{\rm n+1,\beta}} \leq    1\quad\forall\;T\in{\cal T}}$
\\
\\
${\tt and}$ & 
$\quad\displaystyle{H_{\rm n+1,\alpha}=H_{\rm n+1,0}\quad {\rm if}\quad \alpha\in{\cal N}_{0}}$
\end{tabular}
}}}}

\vspace{0.5cm}

Above, we have used
\[
I_{\alpha \beta}\equiv 
\int_{T} \! \lambda_{\alpha}\lambda_{\beta} \; ,
\]
\[
G_{\alpha \beta}\equiv{\bf grad}\lambda_{\alpha}\cdot{\bf grad}\lambda_{\beta}
\]
and ${\cal N}_{0}$ for the subset of boundary nodes, where the magnetic field is determined by the boundary conditions.

Some remarks seem to be convenient before discussing the actual implementation of the numerical algorithm stated above. Firstly, we want to emphasize that the method is flexible enough so as to incorporate any critical current restriction. Thus, anisotropy may be modelled by redefinition of the gradient matrix as
\[
G_{\alpha \beta}= 
(b_{\alpha}b_{\beta}+\gamma^{2}c_{\alpha}c_{\beta})/(4A_{T}^2) \; ,
\]
with $\gamma$ the critical current anisotropy constant (in this case $J_{cx}=J_{cy}/\gamma$). Spatial inhomogeneities in $J_{c}$ may be dealt with, just by using a {\em triangle dependent} scaling factor in $G_{\alpha \beta}$. Finally, field dependent critical currents would be introduced multiplying $G_{\alpha \beta}$ by some definite function of $H$ at the triangle midpoint (barycenter).

From the practical point of view, one will be typically facing a {\em large nonlinear and nonlinearly constrained} optimization problem. Here, by large it is meant that the number of variables, as well as the number of restrictions could be of several thousands. Computationally, this seems quite a formidable task for a small machine. Fortunately, this kind of problem has been met in the modelling of many physical and scientific phenomena (both nature and society tend to optimize and seldom behave as linear systems), and one can find very efficient software solutions, even for a personal computer.

Next, we describe the actual implementation of the previous algorithm which we have adopted for the simulations of several physical problems in this work. It is based on the computing language MATLAB\cite{matlab} and on the fortran package LANCELOT\cite{lancelot} for nonlinear optimization problems. We have found that this combination provides a rather convenient environment, as one can easily handle the finite element mesh and transfer the data structures from one program to the other. The selection of the package LANCELOT  as opposed to the straightforward use of the MATLAB optimization facilities, relates to the higher efficiency of the calculation. Our program is organized according to the following steps. 
\begin{itemize}
\item[(I)] Definition of the geometry of the region $\Omega$.

\item[(II)] Calculation of the finite element triangular mesh on $\Omega$.

\item[(III)] Evaluation of the matrix elements $G_{\alpha\beta}$ and $I_{\alpha\beta}$ for the actual mesh.

\item[(IV)] Set up of the standard input file for LANCELOT on the basis of the previous field profile $H_{\rm n}(\vec{x})$ and the matrix elements calculated in (III)

\item[(V)] Solution of the optimization problem: get the new profile $H_{\rm n+1}(\vec{x})$.

\item[(VI)] Update field profile and iterate from (IV).

\end{itemize}

 It mus be emphasized that the previous process is performed over the full mesh (i.e.: for unknowns $H_{\rm n+1,k}\; ,\;\forall\;k\in{\cal N}$ and under the constraints $|\vec{u}_{T}|\leq 1\;\forall\;T\in{\cal T}$). The definition of regions with or without flux variation ($H_{\rm n+1}(\vec{x})-H_{\rm n}(\vec{x})\neq 0$ or $H_{\rm n+1}(\vec{x})-H_{\rm n}(\vec{x})= 0$), the boundaries between them ($\Gamma$), and the critical or subcritical character ($|\vec{u}|=1$ or $0$) arise from the optimization process itself. That is to say, the totality of mesh points are considered for each iteration, and the optimality process determines where to introduce changes or not. This mathematical simulation is just a replica of the physical laws ruling out the behavior of the superconductor. Local changes are such that global field variations keep minimum, while also fulfilling the minimum global entropy production rate (infinite dissipation if $J > J_c$), and the externally imposed boundary conditions.

\section{TOPOLOGICAL EFFECTS}
\label{seccase}
A rich phenomenology arises in the magnetic field penetration pattern when one considers the influence of the sample's cross section topology. In this part we report on how to understand the concept of critical state for arbitrarily shaped samples. First, we consider the influence of the sample's boundary for the so-defined {\em uniform superconductors} (the current density restriction is homogeneous in space). Secondly, we will concentrate on {\em nonuniform superconductors}, i.e.: the limitations on the current density may change within the sample.

%---------------------------------------------------
\subsection{Uniform superconductors}
%---------------------------------------------------
\begin{figure}
\centerline{
}
\caption{Permanent current density streamlines, obtained for an arbitrarily shaped sample by a cycling the external magnetic field.}  
\label{fig:current}
\end{figure}
%
%--------------------------------------------------
\subsubsection{General features.}
%--------------------------------------------------
%

\begin{figure}
\centerline{
}
\caption{Field penetration profiles within the superconductor produced by ramping up and down of the external field. The vertical axis displays the magnitude of $H$ vs. the shape of the sample (horizontal plane). The external field has been first increased from zero, then decreased and increased again.}  
\label{fig:hysteresis}
\end{figure}

The main property which can be assessed by application of our optimization technique is that, regardless the shape of the cross section, the flux front penetrates through equally spaced contours, which also represent the permanent current streamlines. Such a structure is a straightforward consequence of Eq.(\ref{eq:cece}), as the model implies a constant gradient, and it has been illustrated in Fig.\ref{fig:current}. Two regions are defined by the clockwise or anti-clockwise circulation of the current, as corresponds to a magnetic history in which the applied field has been increased and then decreased. Notice the boundary line in between.

\begin{figure}
\centerline{
\includegraphics[height=3.in]{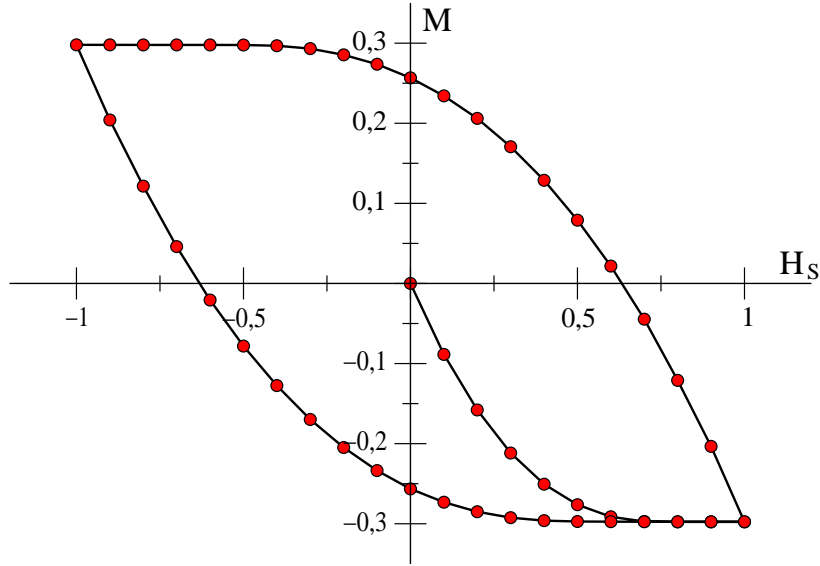}
}
\caption{Magnetic hysteresis loop corresponding to the process illustrated in Fig.\ref{fig:hysteresis}.}  
\label{fig:mh}
\end{figure}

Magnetic hysteresis has also been simulated, and one can observe the standard {\em Rayleigh-like} diamagnetic loop already predicted by C.P. Bean for constant $J_c$ in infinite slabs or circular cylinders. In fact, the flux penetration and exit that takes place along the process (see Fig.\ref{fig:hysteresis} and Fig.\ref{fig:mh}) are nothing but the generalization of the behavior for ideal geometries. The external drive pulls up and down across the boundary, while maximum gradient profiles are defined within the sample. The interface (free boundary) between active and passive regions (as regards flux changes) always corresponds to a line of constant distance to the sample's perimeter.

%---------------------------------------------------
\subsubsection{The effect of corners.}
%---------------------------------------------------
%
A number of outstanding effects related to the penetration of magnetic flux in superconductors with corners have been reproduced within the framework of our model. We will concentrate on the field profiles for the cross section displayed in Figs.\ref{fig:freebound} and \ref{fig:paths}. This example simultaneously holds concave and convex corners, and this enables to visualize several pecularities for non ideal geometries. 

Fig.\ref{fig:gamma} shows a sequence of increasing field penetration profiles. Ten contours are plotted, starting with the applied field at the surface. Notice that, as expected, equally spaced curves are obtained. However, the particular shape of the outer perimeter leads to a splitting process, which takes place as soon as the {\em nearest} points of the penetrating boundaries contact each other. Further profiles are characterized by a disconnected structure in which the free boundary between flux penetrated and flux free regions is the union of disjoint curves ($\Gamma = \Gamma_{1}\bigcup\Gamma_{2}$).

Another group of properties is related to the corners themselves. For the case of convex corners, we have observed the formation of the so-called {\em d-lines}\cite{schuster} as the locus of points where the current paths display sharp bends. Such lines are intuited in Fig.\ref{fig:gamma}, and have been outlined in Fig.\ref{fig:paths}. Notice that the flux channels do never cross such lines. Geometrically, as vortices always move perpendicular to the current streamlines, they have to be stopped at the points where $\vec{J}$ abruptly changes direction. 

For concave corners, a quite different phenomenon is observed. As it was already reported for the case of rectangular geometry\cite{schuster}, concave corners become a source of magnetic flux in the critical state. This fact is apparent in our simulations. Notice that flux channels display divergence at the concavities, because the flux which replenishes a finite region must be supplied from a single point. Such behavior is of high importance because this point is a natural candidate for the triggering of magnetic instabilities. Just recall that the induced electric field at any point may be identified with the amount of magnetic flux which has crossed the unit length measured along $\vec{E}$ in the unit time. Thus, $\vec{E}$ would become infinite for an ideal corner and, in any case, large for a real one. As the amount of generated power per unit volume in the critical state is $EJ_{c}$, heat removal would first fail at such points (provided the sample is thermally homogeneous).
\begin{figure}
\centerline{
}
\caption{Splitting of the free boundary in the field penetration process for a sample with corners.}  
\label{fig:gamma}
\end{figure}
%
%---------------------------------------------------
\subsection{Nonuniform superconductors}
%---------------------------------------------------
In this section, we present numerical simulations for samples displaying nonhomogeneous behavior of $J_{c}$ across the section. In principle, any physically meaningful dependence $J_{c}(x,y)$ could be dealt with. Here, we concentrate on two specific cases: granular samples and samples with voids.
%---------------------------------------------------
\subsubsection{Granular samples}
%---------------------------------------------------
\begin{figure}
\centerline{
}
\caption{Penetration of magnetic field in a granular sample. The first panel illustrates the critical current inhomogeneity (superconducting granules within a poorly superconducting matrix). A sequence of field penetration contours follows.}  
\label{fig:granular}
\end{figure}
Granularity has been a main fact in the development of superconductivity, mainly since the venue of the high temperature superconducting oxides. Owing to their complex chemical structure, these materials have been widespread obtained in ceramic form, typically as a collection of superconducting grains embedded in a poorly superconducting matrix. Their properties have been usually described at a mean field level, characterizing the matrix with some effective parameters\cite{db}. However, either analytical or semianalytical statistical theories have been developed under certain simplifying assumptions. Here we emphasize that the magnetic behavior for any geometrical arrangement may be simulated with our numerical approach. The basic idea that magnetic flux is first screened from the whole volume of the sample by means of {\em intergranular currents} and further penetrates within a collection of basically disconnected grains is visualized.

Fig.\ref{fig:granular} depicts the successive field penetration profiles for a granular sample with rectangular cross section. A collection of cylindrical granules with elliptical cross sections has been used, assuming a constant critical current within them ($J_{c{g}}$) which is ten times larger than the intergranular limit ($J_{c{I}}=J_{c{g}}/10$). Notice that, at first, a critical state characterized by $J_{c{I}}$ is built for the intergranular region. The superconducting grains are {\em connected} in the sense that the field is excluded from the region in between. As soon as the intergranular shielding is not intense enough, a {\em double} critical state is defined. The magnetic flux penetrates with constant slope $J_{c{I}}$ for the intergranular region, and with the slope $J_{c{g}}$ within the grains.

%---------------------------------------------------
\subsubsection{Samples with voids.}
%---------------------------------------------------
%
\begin{figure}
\centerline{
}
\caption{Penetration of magnetic field in a sample with a void. The magnetic field intensity is displayed vertically, grounded on the meshed structure used for the simulation.}  
\label{fig:void}
\end{figure}
We have simulated the field penetration profiles within a long superconductor of square cross section with a long cavity of circular shape along the axis, and close to one of the edges (see Fig.\ref{fig:void}). The structure of the current distribution of such a defect was already discussed by Campbell \& Evetts\cite{campbellevetts} and has been accurately reproduced by our optimization principle. Two different simulation methods have been used, producing the same results for the magnetic field evolution.

On the one side, one can describe the whole region as a nonhomogeneous superconductor, formed by a circular cylinder with zero critical current $J_{c,h}=0$ ($\Omega_{h}$ will be used for this region), embedded in a matrix with finite critical current $J_{c,m}\neq 0$ ($\Omega_{m}$ will be used for this region). Then, one has $\Omega =\Omega_{h}\bigcup\Omega_{m}$ and the set of boundary nodes ${\cal N}_{0}$ corresponds to the outer square perimeter $\partial_{\rm O}\Omega_{m}$. Minimization is performed under the restrictions
\begin{eqnarray}
\sum_{\alpha ,\,\beta\in{T}}
{H_{\rm n+1,\alpha}}G_{\alpha \beta}{H_{\rm n+1,\beta}}
\left\{
\begin{array}{ll}
= 0 & \forall\quad T \in {\cal T}(\Omega_{h})\\
\leq 1 & \forall\quad T \in {\cal T}(\Omega_{m})
\nonumber
\end{array}
\right.
\\
\nonumber
\\
H_{\rm n+1,\alpha}=H_{\rm n+1,0}\quad {\rm if}\quad \alpha\in\partial_{\rm O}\Omega_{m} \; .
\nonumber
\end{eqnarray}
Notice that equality constraints correspond to the absence of a pinning structure for the magnetic flux within the hole.

On the other hand, one can also solve the problem by excluding the hole from the mesh. Then, the boundary of the superconductor splits into an outer boundary $\partial_{\rm O}\Omega_{m}$ (square perimeter) and an inner boundary $\partial_{\rm I}\Omega_{m}$ (circular perimeter). Minimization is performed according to
\begin{eqnarray}
\sum_{\alpha ,\,\beta\in{T}}
{H_{\rm n+1,\alpha}}G_{\alpha \beta}{H_{\rm n+1,\beta}}
\leq 1 & \quad\forall\quad T 
\nonumber\\
\nonumber\\
H_{\rm n+1,\alpha}=
H_{\rm n+1,0}&\quad{\rm if}\quad \alpha\in\partial_{\rm O}\Omega_{m}
\nonumber\\
\nonumber\\
H_{\rm n+1,\alpha}=H_{\rm n+1,\beta}&\quad \forall \quad\alpha , \beta\in \partial_{\rm I}\Omega_{m}  \;.
\nonumber
\end{eqnarray}
This means that the magnetic field along the outer boundary is determined by the value of the external excitation, and that it must be constant along the boundary of the hole. This last condition stems from the fact that the current must flow parallel to the surface of the hole. Then ${\bf grad}H_{\rm n+1}=\vec{u}$ is perpendicular to this surface, and thus $H_{\rm n+1}(\partial_{\rm I}\Omega_{m})=constant$. Notice that this constant is {\em a priori} unknown and will be determined by the optimization process itself. 

As a further technical detail, before discussing the physics of the problem, we would like to comment that the accuracy of the solution may be increased within a reasonable computational time, by a specific subdomain decomposition related to the structure of the solution itself. Thus, in Fig.\ref{fig:void} we display the actual decomposition of $\Omega_{m}$ in four triangular subregions, according to the well known {\em d-line} structure of the square cross section, in which sharp bends of the current lines appear along the diagonals. Of course, the optimal solution itself should not depend on the specific meshing process, but a much larger number of triangles is required for other choices.

\begin{figure}
\centerline{
\includegraphics[height=4in]{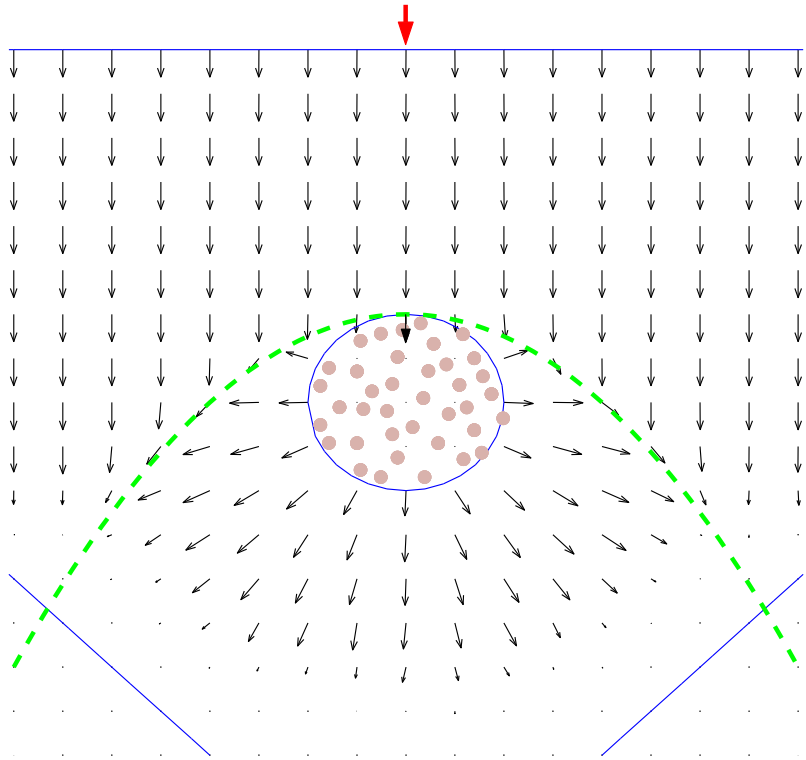}
}
\caption{Flux channels in the vicinity of a hole, as represented by a vector plot of $-\vec{u}$, proportional to the magnetic field current density $-\vec{J}_{H}$ (see Eq.\ref{eq:continuity} and aside). A collection of vortex lines is depicted within the void.}  
\label{fig:fluxlines}
\end{figure}

Let us now analyze the physical process of magnetizing such a sample. As it is apparent from Fig.\ref{fig:void}, the magnetic field levels display a constant gradient within the sample. $H$ linearly decreases from a constant value at the surface (the applied field $H_{S}$ which equals $H_{\rm n+1,0}$ for the n{\em th} time step), displaying a characteristic {\em plateau} at the hole area. This plateau changes height along the process, as it is connected to the surface by a constant slope region. Thus, the magnetic field within the sample is {\em refresehed} in the following way. At the beginning, flux penetrates along channels, which are perpendicular to the outer perimeter and finish at an equidistant square front. Then, as soon as this front reaches the hole a new physical mechanism begins. This has been illustrated in Fig.\ref{fig:fluxlines} by means of a vector plot of the field $-\vec{u}=-{\bf grad}H_{\rm n+1}$ at a particular time step. Recall that, according to the optimal control equations, the magnetic field current density $\vec{J}_{H}$ is proportional to the vector $-\vec{u}\;$ (see Eq.\ref{eq:continuity} and aside). Then, it is apparent that, while following the indicated paths, flux lines will preferentially penetrate along the highlighted channel. This channel carries all the flux which, on the one side replenishes the volume of the hole, and in addition {\em diverges} radially through its surface, penetrating again into the superconductor. We want to emphasize that, as remarked by other authors\cite{campbellevetts,schuster2}, the magnetic flux outcoming from the hole meets the channels coming from the outside boundary in a parabolic {\em d-line}. This is a simple consequence of the constant critical current ansatz. Discontinuity arises by sharp bending field contours, and this takes place at points equidistant to the center of the hole and to a line parallel to the horizontal edge of the sample. Trivially, such points are the locus of the parabola $y=D+R-x^{2}/4R$, with $D$ standing for the vertical coordinate of the center of the circle, and $R$ for its radius (the origin has been taken at the center of the square). Notice the nice fit of this curve and the place where $\vec{u}$ changes direction in our simulation.

\section{CONCLUDING REMARKS}
\label{secdiscuss}
We have applied a direct numerical method for investigating several topological properties of the critical state in type-II superconductors. The variational interpretation of Faraday's law, constrained by the depinning threshold in the form of the rule $\vec{J}\in\Delta$ for the current density, leads to a standard nonlinear optimization problem on the finite element discretization. This has been solved by using standard algorithms in the case of long cylinders with arbitrary cross section, and the isotropic case ($\Delta$ is a circle). Essentially, the values of the magnetic field at the mesh points ($H_{\rm n+1,k}$) are treated as unknowns for each time layer and a minimizer for the global change $\int_{\Omega}\! | H_{\rm n+1} - H_{\rm n} |^{2}$ is sought. As a result of this method, we obtain the free boundary between the regions (un)affected by flux changes, and the circulating current density, which are outputs of the theory. The algorithm itself chooses whether $H_{\rm n+1,k}-H_{\rm n,k}$ is zero or not and reproduces the critical behavior $|\vec{J}|=J_{c},0$.

The high flexibility of the theory allows to simulate a number of physical phenomena as the influence of arbitrary corners in flux penetration, the splitting of free boundaries, spatial inhomogeneities in the critical current, and the calculation of critical state properties in multiply connected topologies (samples with voids).

The fundamental issue $|\vec{J}|=J_{c},0$ of our numerical studies for the 2D problem confirms the result: ${\bf grad}H^{\,*}/J_{c}=\vec{u}^{\,*}=\vec{p}^{\,*}/p^{\,*}$, which is a straightforward consequence of the optimal control formulation for the variational statement. This reflects a very general property ($\vec{u}^{\,*}\in\partial_{0}\Delta_{\perp}$), relating to the fact that the cost function does not explicitely depend on the control variables. In particular, it can be shown that the still challenging full 3D problem is also characterized by the condition $|\vec{J}|=J_{c},0$. This is shown below. The 3D critical state problem may be posed as 

\vspace{0.5cm}

{{\hspace{-.25cm}\fboxsep
1.ex
%\fbox
{\parbox{0.9\textwidth}{ 
\begin{tabular}{rl}
${\tt Minimize}$ & $ {\cal C}[\vec{H}_{\rm n+1}(\vec{x})]
={\displaystyle \int}_{{\relax{\rm I\kern-.18 em R}}^3}\; | \vec{H}_{\rm n+1} -
\vec{H}_{\rm n} |^{2}$
\\
\\
${\tt for}$ & $ \nabla\times\vec{H}_{\rm n+1}=\vec{J}_{e}\quad{\rm in}\quad{\relax{\rm I\kern-.18 em R}}^3{\setminus}\Omega\quad{\rm and}\quad\nabla\times\vec{H}_{\rm n+1}\in\Delta\quad{\rm in}\quad\Omega$
\end{tabular}
}}}}

\vspace{0.5cm}

Now, let us assume that the restriction set $\Delta$ is convex (this is not necessary, but simplifies the analysis). Then, the optimal control solution $\nabla\times\vec{H}_{\rm n+1}^{\,*}\equiv\vec{u}^{\,*}$ verifies
\[
{\cal C}(\vec{u}^{\,*})\leq {\cal C}[\,(1-\theta )\vec{u}^{\,*}+\theta \vec{u}\,] \quad\forall \; \vec{u}\in\Delta,\;\; 0\leq\theta\leq 1 \; ,
\]
and this leads to
\[
\lim_{\theta\to 0}\,\frac{1}{\theta}\,[\,{\cal C}(\vec{u}^{\,*}+\theta(\vec{u}-\vec{u}^{\,*}))-{\cal C}(\vec{u}^{\,*})\,]\geq 0 \quad\forall \; \vec{u}\in\Delta\; ,
\]
which one can identify as the condition that the {\em directional derivative} for allowed variations within $\Delta$ is nonnegative around the minimum.

If one introduces the actual functional of our problem, the following condition is met
\[
\int_{{\relax{\rm I\kern-.18 em R}}^3 }\; (\vec{H}_{\rm n+1}(\vec{u}^{\,*})-\vec{H}_{\rm n})\cdot(\vec{H}_{\rm n+1}(\vec{u}\,)-\vec{H}_{\rm n+1}(\vec{u}^{\,*})) \geq 0 \quad \forall \;\vec{u}\in\Delta \;.
\]
Then, assuming that a vector field $\vec{e}$ exists on ${\relax{\rm I\kern-.18 em R}}^3$ such that $\vec{H}_{\rm n+1}(\vec{u}^{\,*})-\vec{H}_{\rm n}\equiv -\nabla\times\vec{e}\;$, one has
\[
\int_{{\relax{\rm I\kern-.18 em R}}^3 }\;  (\nabla\times\vec{e}\,)\cdot[\vec{H}_{\rm n+1}(\vec{u}^{\,*})-\vec{H}_{\rm n+1}(\vec{u}\,)]\geq 0 \quad \forall \; \vec{u}\in\Delta \;.
\]
On using vector formulas, and assuming that the magnetic field at large distances is determined by the external sources, the integral may be transformed to
\[
\int_{{\relax{\rm I\kern-.18 em R}}^3}\;\vec{e}\cdot(\vec{u}^{\,*}-\vec{u}\,)
\geq 0\quad \forall \; \vec{u}\in\Delta\;.
\]
Eventually, as any candidate control function should verify the external current restriction, we get
\[
\int_{\Omega}\; \vec{e}\cdot(\vec{u}^{\,*}-\vec{u}\,)  \geq 0 \quad \forall\; \vec{u}\in\Delta \;.
\]
In the case of pointwise controls (the critical current at the point (x,y,z) does not depend on sample properties at others) this is equivalent to\cite{lions}
\[
\vec{e}\cdot\vec{u}^{\,*} \geq \vec{e}\cdot\vec{u}  \quad \forall\; \vec{u}\in\Delta 
\quad \Longrightarrow\quad  \vec{u}^{\,*}\in\partial_{0}\Delta
\;,
\]
that is to say, the {\em maximum projection} rule implies that the control lies on the boundary (in the generalized sense, which admits the trivial solution $\vec{u}^{\,*}=0$ if allowed by the boundary conditions). This ensures that the rule $|\vec{J}|=J_{c},0$ is valid for the general 3D problem within the isotropic case.

From the numerical side, the solution of the 3D problem needs a more encompassing view point and more sophisticated (though available) mathematical tools. On the one side, the finite element mesh must be a tessellation by volumes of some shape. A tetrahedral mesh would seem a natural generalization. Then, one can make use of tetrahedral nodal functions $\lambda_{i}$, which verify analogous properties to their 2D counterpart\cite{bossavit}. In fact, a vectorial built (the so-called {\em Whitney elements}) is typically used in computational electromagnetism as the right geometrical framework. For instance, the natural basis for the magnetic field ($\vec{H}=\sum{H_{e}\vec{w}_{e}}$) is obtained from $\vec{w}_{e}\equiv\lambda_{k}\,{\bf grad}\lambda_{l}-\lambda_{l}\,{\bf grad}\lambda_{k}$, where $e$ is used for the edge joining nodes $k$ and $l$. Finally, although the current density restriction operates on the finite region $\Omega$, one must minimize over the whole space ${\relax{\rm I\kern-.18 em R}}^3$. This process requires special techniques, such as the restriction to some artificial boundary, the use of adaptive meshing, with larger and larger elements for distant points, etc. Further development of the theory is planned along this line.

\section*{ACKNOWLEDGEMENTS}
Financial support from the Research Program of the University of Zaragoza (Project no. 224-92) is acknowledged.

\vspace{1.5cm}
%\hline

\footnotetext{Contributed chapter in {\em Horizons in Superconductivity Research}, Nova Science Publishers, 2003}

\end{document}